
\input phyzzx
\input tables
\def\half{{1\over 2}}
\def\dbra{\langle\!\langle}
\def\dket{\rangle\!\rangle}
\titlepage
\title{A Model Study of the Strength Distribution for a Collective
       State  Coupled with Chaotic Background System}
\vskip .5cm
\centerline{Hirokazu AIBA~ and ~Toru SUZUKI$^*$}
\vskip .3cm
\address{Koka Women's College, 38 Kadono-cho Nisikyogoku, Ukyo-ku, 615 Kyoto,
Japan}
\address{$^*$Research Center for Nuclear Physics, Osaka University, 567
Ibaraki, Japan}
\vskip .5cm
\abstract{
We consider a model in which a collective state couples to
a large number of background
states. The background states can be chosen to have properties which
are classically characterized as regular or chaotic. We found that
the dynamical nature of the background system considerably
affects some fluctuation properties of the strength function. }
\endpage
\centerline{\bf 1. Introduction}

Over the last twenty years a number of new giant resonances
have been found in a broad
range of the nuclear table.  They are coherent particle-hole
excitations carrying a large fraction of the energy weighted
sum rules.  In most cases they are embedded in a continuum and
are damped due to particle escape.  At not much high excitation
energies, however, the damping of giant resonances is dominated
by spreading, i.e., by  coupling to a huge number of background
states. It is also known that there is in some cases a mechanism
which reduces the width of resonances [1].

In many theoretical calculations the spreading of a collective
particle-hole state
is accounted for by coupling to more complicated states, e.g.,
two-particle two-hole states. As the number of these states in
heavy nuclei becomes enormously large it is necessary to truncate
the configuration space, or to introduce some kind of
statistical treatment.
For instance, in ref.2, the spreading was discussed by means of the
random matrix theory. Aside from the reduction mechanism mentioned
above, the coupling matrix element to each of the background states
is expected to behave randomly, reflecting a random nature of the
latter. In fact, the single-particle width distribution of
neutron resonances obeys the Porter-Thomas distribution consistent
with the random matrix theory [3]. We may note that  level
statistics of excited states have been investigated
in realistic shell model calculations[4,5].
Not much is known, however, if or
what properties of a strength distribution of collective
state are related to the dynamical properties
of the background states.

In the present note we consider a model which imitates a
coupling of a  collective state to  background states, the latter
being chosen to have chaotic or regular dynamical properties.  In
this manner we investigate wheter there is a characteristic
signature in the strength function which reflects a nature
of the background system.
\vskip 1cm
\centerline{\bf 2. Model}

In the study of giant resonances, the Hamiltonian is normally
diagonalized within the truncated
space of one-particle one-hole states plus background states,
mainly two-particle two-hole states.
In the present model we replace the background states
 with  eigenstates of a system whose classical counterpart
is well studied, while a collective mode is represented by
a single boson excitation independent of the background system.
The structure of the collective strength function
depends both on the coupling Hamiltonian of the   collective
mode to the background and also on the dynamics of the
background system itself.
We adopt a simple ansatz for the coupling in order to
focus on  the latter effect.
We start with the Hamiltonian
$$\eqalignno{
       H &= H_{\rm coll} + H_{\rm bg} + V_{\rm coupl}, &(1) \cr
       H_{\rm coll} &= \epsilon B^\dagger B, \quad
       H_{\rm bg} = \half (p_x^2+p_y^2+x^4+y^4) - kx^2y^2. &(2) \cr}
$$
Here $H_{\rm coll}$ represents a  Hamiltonian for the
collective excitation, where
$B^\dagger$ and $B$ represent boson creation and annihilation
operators. We write the boson vacuum as $|0)$. As the
Hamiltonian $H_{\rm bg}$ of the background system we choose
that of a two-dimensional anharmonic oscillator characterized
by a single parameter $k$. The classical counterpart of this
Hamiltonian describes a transition from an integrable to a chaotic
dynamical system[6]. The quantum spectra and the wave
function characteristics
follow the same trend[7,8]: At $k=0$ the system is separable while
at $k=0.6$, for instance, the nearest neighbor spacing distribution
of the quantum levels shows the Wigner distribution which is typical
for chaotic systems.
We denote the eigenstate of $H_{\rm bg}$ as $|n\dket$ (
$|n=0\dket$ for the ground state).
In the diagonalization of $H_{\rm bg}$, we take as the basis
states of the background system the eigenstates
of an uncoupled harmonic oscillator whose frequency $\Omega$ is
determined so as to optimize the  diagonalization of $H_{\rm bg}$ [7,8].
They are denoted by $\ket{N}$ ($\ket{N=0}$
for the ground state) where $N$ stands for a pair of integers,
i.e., numbers of oscillator quanta in the $x-$ and the
$y-$directions.

The interaction $V_{\rm coupl}$ represents the coupling between the collective
state
and the background states.
We take a simple ansatz for the coupling,
$$
         V_{\rm coupl}=\chi \sum_{N\ne 0}
                       (B^\dagger \ket{0}\bra{N} + {\rm h.c.})    \eqno(3)
$$
characterized by the strength parameter $\chi$.
Note that the state $\ket{N}$ is different for each value of $k$
because of the difference in the $\Omega$ value.
Due to the simple form of eq.(3),   regular or random
behavior of the coupling matrix elements depends entirely on the
dynamics of the background system which can be controlled
by the parameter $k$.  The assumption
(3) may not be  realistic, as the coupling
strength of the actual nuclear system would decrease for complicated
states (i.e., large $N$).
We expect, though, eq.(3) is sufficient for the present purpose of
studying a qualitative difference coming from the dynamical
structure of the background system.

We divide the whole space into four parts:

\item{1.} The ground state of $H_{\rm coll} + H_{\rm bg}$:
          $\,\ket{0;0}\equiv |0)|0\dket$.
\item{2.} The one collective boson state: \quad
          $\ket{1;0}\equiv B^\dagger|0)|0\dket$.
\item{3.} The background states: \qquad\qquad
          $\,\ket{0;n}\equiv |0)|n(\ne 0)\dket$.
\item{4.} Other states.

\noindent
In diagonalizing the Hamiltonian (1), we neglect
the coupling between space 1 and space 2, and also omit
the space 4. These approximations produce negligible effect.

The eigenstates of $H_{\rm bg}$ are classified into several
symmetry classes with no coupling among them [7]. In the present
calculation we consider only those states which belong to the
class symmetric in the $x-, y-$ and the diagonal($x=y$)-directions.
First $H_{\rm bg}$ is diagonalized within a large space (the
number $N_{\rm max}$ of basis states is
5776) and then the lowest 800 states are included in the
diagonalization of the total Hamiltonian. As for the values of
$k$, we consider three typical cases, i.e., $k=$0.0 for an
integrable background system, 0.2 for a partially
irregular system, and 0.6 for an almost chaotic system. The value of
$\epsilon$ has been fixed to 220, so that the collective state is
located in the middle of the background 800 states, and thus
a large number
of background states can be found in the neighborhood.  For a fair
comparison of the role of integrable versus chaotic background system
one might better adjust $\epsilon$ value
for each $k$ to give a similar local  level
density. Although the
asymmetry in the level density of the background states around
$\epsilon$ does affect the third moment of the strength
function as discussed later, its effect on the fluctuation properties
of the strength distribution is expected to be small. The coupling
strength is mostly fixed to $\chi=1.0$ in the calculation below.
\vskip 1cm
\centerline{\bf 3. Distribution of Strength}

Collective strengths are measured with respect to
the operator $\hat{O}=B^\dagger + B$, i.e.,
$$
        S(E) = \sum_{\alpha}\delta(E-(E_\alpha-E_{g.s.}))
               |\,_{\rm tot}\!\bra{\alpha}\hat{O}\ket{g.s.}_{\rm tot}|^2,
     \eqno(4)
$$
where $\ket{\alpha}_{\rm tot}$ denotes an eigenstate of $H$,
 and $E_\alpha$ the corresponding eigenvalue.

Before discussing the strength function $S(E)$ in detail, we
first examine the distribution of the coupling matrix elements
$v_n\equiv\bra{1;0}V_{\rm coupl}\ket{0;n}$.
According to our choice of the interaction it is expected that the
coupling matrix elements $v_n$ would behave
regularly for $k$ small and randomly for $k$ in the chaotic regime.
This is indeed so as seen in Fig.1 where the distribution of the
coupling matrix elements is shown.
\midinsert
\begintable
Figure 1
\endtable
\endinsert
\noindent
It is seen that the matrix element values for $k=$0.0 are
concentrated at $\pm 1$, while for
$k=0.6$ they show almost the gaussian distribution centered at zero
and with the width of around 1.1. In fact, for a variable $v_n$
composed of a large number of random elements
the distribution would be a gaussian,
$$
     P(v_n) = {1\over\sqrt{2\pi a_n}}\exp (-{v_n^2\over 2 a_n^2}), \eqno(5)
$$
with $a_n=1$ from the normalization of $v_n$.
\midinsert
\begintable
Figure 2
\endtable
\endinsert
%
Figure 2 shows a strengh function $S(E)$.
In spite of the difference in the coupling matrix elements as
seen above, the shape of $S(E)$ looks rather
similar for $k$=0.0 and 0.6.  For the sake of quantitative
discussion of the gross structure, let us consider
the cumulant $\VEV{E^n}_c$ of the strength function.
Several low order cumulants are given by
$$
    \VEV{E}_c  =\VEV{E}, \quad \VEV{E^2}_c=\VEV{E^2}-\VEV{E}^2, \quad
    \VEV{E^3}_c  =\VEV{E^3}-3\VEV{E^2}\VEV{E}+2\VEV{E}^3, \eqno(6)
$$
where $\VEV{E^n}$ is the $n$-th energy-moment of the strength function defined
as,
$$
     \VEV{E^n}\equiv \int E^nS(E)dE.             \eqno(7)
$$
We show in Table 1 the calculated cumulants up to the third order at
three $k$ values and for $\chi =1.0$. Let us consider the
second cumulant which can be written as,
$$\eqalign{
      \VEV{E^2}_c &=\sum_{n(\ne 0)}v^2_n, \cr
     v_n^2 &=\chi^2|\bra{0} 0\dket|^2\sum_{N,N'(\ne 0)}
      \bra{N}n\dket\dbra n\ket{N'}.          \cr  }\eqno(8)
$$

\midinsert
\begintable
Table 1
\endtable
\endinsert

\noindent
Eq.(8) shows that, if we do not truncate the number of $|n\dket$, we
would obtain almost the same values of $\VEV{E^2}_c$ determined by
$N_{\rm max}$ for any $k$ [9].
Indeed, the actual values of $\VEV{E^2}_c$, as listed in Table 1,
are all close to 800 for the three values of $k$.
The third cumulant can be written as,
$$
      \VEV{E^3}_c=\sum_{n(\ne 0)}(\omega_n-\epsilon)v^2_n,  \eqno(9)
$$
where $\omega_n$ denotes the eigenvalue of $H_{\rm bg}$.
{}From Table 1, we find that the value of $\VEV{E^3}_c$ decreases as
$k$ increases. This trend is mostly due to the difference in the level
densities around $\epsilon$ as mentioned before: If we artificially
use the same sequence of $\omega_n$ for the three $k$ values we
obtain almost the same values for the third cumulant.
The apparent similarity of the gross structure of the strength
function may be understood in this manner. We mention that the
shape of the strengh function is not much different for
different values of $\chi$, except that the width of the
distribution is scaled accordingly.
\midinsert
\begintable
Figure 3
\endtable
\endinsert
%


The similarity of the strengh function is only superficial,
however. This is seen in the strength distribution $P(S)$ as given
in the upper part of Fig.3. The smooth curve shows a Porter-Thomas
distribution which is expected to hold in the chaotic system as
given in the random matrix theory [3,4]. Difference due to the
dynamics of the background system can be more clearly seen in
the lower part of Fig.3, where the distribution of the
amplitude $\sqrt{S}$ corrected for the energy dependent
factor $((E-\epsilon)^2+(\Gamma/2)^2)^{1/2}$ is shown. The latter
has been introduced to remove approximately the energy denominator
contribution:
If we assume constant coupling matrix elements $v_n=v_c$
 and an equal level distance $D_c$,
the strength function will be given by
$S(E)\simeq (\Gamma/2\pi)/\{(E-\epsilon)^2+(\Gamma/2)^2\}$, where
$\Gamma=2\pi v_c^2/D_c$ [9]. This may be contrasted to the
present calculation where the mean square value of $v_n$ gives
$\bar{v^2_n}\approx\chi^2$(see Table 1) and the mean
level distance is $\bar{D}\approx 0.5$. Thus we choose
$\Gamma=4\pi\chi$ in Fig.3.
 The result shows that at $k=0.6$ the distribution
follows a gaussian, while at $k=0.0$ it is peaked at unity.
Notice that the former distribution is generic, while the latter,
corresponding to the  regular background system,
depends on the specific structure of the coupling.


The strength distribution is one of the signatures
which characterize the underlying dynamics of the system. It
reflects only a part of the structure in the strength function
of Fig.2. For instance,  the structure
of the eigenstates of $H_{\rm bg}$ as a function of energy
changes regularly at small $k$ values, while at $k=0.6$
it strongly fluctuates from state to state[8]. A similar behavior
may be expected to occur in the present coupled system. For this purpose
we perform a moment analysis similar to the one applied for a
multifractal system[10]. This analysis takes into account
some features of the energy-strength correlation, and therefore
can be another characteristic measure of the strength function
independent of the distribution $P(S)$.
 Here the whole energy interval $\Delta E$ is divided
into $L$ segments each having a width  $\delta E=\Delta E/L$.
(We take $\Delta E=E_{800{\rm th}}-E_{g.s.}$ in the present case.)
The strength in each segment is summed up to
give the strength $P_j (j=1,\ldots ,L)$ for the $j-$th segment,
with the normalization
condition $\sum_j P_j =1$. The $m-$th moment at the scale
$\delta E$ is defined by
$$
          M_m(\delta E) \equiv \sum_{j=1}^L P_j^m.             \eqno(10)
$$
We then study its behavior  as we refine the
scale, e.g., as $L=2 \rightarrow 2^2 \rightarrow 2^3 $ etc.
\midinsert
\begintable
Figure 4
\endtable
\endinsert
Figure 4 shows the dependence of the moments $M_m$ for $m=2$
to 5 on the scale $\delta E$ at $k=0.0$ and 0.6. We also show,
for the sake of comparison, the one for the ideal case of
equidistant eigenvalues with equal strength.
Since we have a discrete spectrum, the
moment $M_m$ for small $\delta E$, i.e., for a large number of
segments eventually reaches a fixed value. It is seen that
the slope becomes almost constant around $\delta E\simeq 2$.
We now consider the fractal dimension $D_m$ defined by
$$
    D_m\equiv\lim_{\delta E\rightarrow 0}{B_m(\delta E)\over m-1},\quad
    B_m(\delta E)={\log M_m\over\log {\delta E}}.    \eqno(11)
$$
In  practice, the expression for $B_m(\delta E)$ is replaced with
the ratio of the difference of $\log M_m(\delta E)$
to that of $\log \delta E$ in the appropriate interval of
$\delta E$. The quantity $D_m$ reflects a state-to-state
fluctuation of the strength function. In Table 2 we
show the calculated values of $D_m$ for $m=2$ to 5.

\midinsert
\begintable
Table 2
\endtable
\endinsert

\noindent
The adopted interval of $\delta E$ is indicated in Fig.4.
The result shows that the strength
function at  $k=$0.6 has a smaller $D_m$ value
than that for $k=$0.0 and for the ideal case. We also see
from Fig.4 that the
dependence of $M_m$ on $\delta E$ is smoother for $k=0.6$ than
the other cases. These results may indicate that the scaled moments
(10) provide another characteristic signature which reflects the
underlying dynamics of the system.
\vskip 1cm
\centerline{\bf 4. Summary}

In summary, we constructed a model which simulates a collective
state coupled to a background. We took a two-dimensional
anharmonic oscillator as the background system which exhibits a regular or
a chaotic spectrum depending on the   parameter $k$.
We found that the difference in the dynamics of
the background system causes
a characteristic difference in the strength distribution.
We also suggested that the scaled moment analysis might be
useful in the study of an energy-strength correlation,
although it requires a further study in order to make clear
whether the conclusion survives in a more generic context.
In the actual study of the strength function of a physical system,
e.g., of the nuclear giant resonances, the finite experimental
resolution as well as the effect of a continuum put a
restriction on the applicability of the method presented
here. Other possible methods, such as the autocorrelation function
and the Fourier transform,
are now under investigation[11].

\par\vskip .3cm
\leftline{\bf References}
\item{[1]}G.F.Bertsch, P.F.Bortignon and R.A.Broglia,
 Rev.Mod.Phys.{\bf 55}(1983)287.
\item{[2]}T.Guhr and H.A.Weidenm\"uller, Ann.Phys.{\bf 193}(1989)472.
\item{[3]}C.E.Porter and R.G.Thomas, Phys.Rev.{\bf 104}(1956)483.
\item{[4]}T.A.Brody et al., Rev.Mod.Phys. {\bf 53}(1981)385.
\item{[5]}M.S.Bae et al., Phys.Rev.Lett.{\bf 69}(1992)2349.
\item{[6]}H.-D. Meyer, J.Chem.Phys.{\bf 84}(1986)3147.
\item{[7]}Th.Zimmermann {\it et al.}, Phys.Rev.{\bf A33}(1986)4334;
\item{   }O.Bohigas, S.Tomsovic and D.Ullmo, Phys.Rep.{\bf 223}(1993)43.
\item{[8]}H.Aiba and T.Suzuki, RCNP 1991 Annual Report and to be published.
\item{[9]}A.Bohr and B.R.Mottelson, {\it Nuclear Structure} Vol.1
         (Benjamin, 1969) Chap.2D.
\item{[10]}J.L.McCauley, Phys.Rep.{\bf 189}(1990)225;
\item{    }see also, S.N.Evangelou and E.N.Economou, Phys.Lett.{\bf
A151}(1990)345.
\item{[11]}H.Aiba, S.Mizutori and T.Suzuki, in preparation.
\endpage
\input tables
\centerline{TABLES}
\vskip 1cm
%
\item{Table~ 1.}Cumulants of the strength function from the
               first to the third order for $\chi=1.0$.

 \midinsert
 \begintable
          $k$ \vb $\VEV{E}_c$ \vb $\VEV{E^2}_c$ \vb $\VEV{E^3}_c$ \cr
  0.0\vb 220.0\vb 842.5\vb 28516.5\cr
  0.2\vb 220.0\vb 779.5\vb 15784.9\cr
  0.6\vb 220.0\vb 965.9\vb -1942.3\endtable
\endinsert
\vskip 2.0cm

\item{Table~ 2.} Fractal dimension $D_m$~
      ($m=2$ to 5)~ at $k=0.0$ and 0.6 for $\chi=1.0$.
       Also those for the ideal case of equidistant energy
       eigenvalues with equal strength are listed for comparison.

 \midinsert
 \begintable
          $k$ \vb $D_2$ \vb $D_3$ \vb $D_4$ \vb $D_5$ \cr
  0.0   \vb 0.73\vb 0.71\vb 0.70\vb 0.69    \cr
  0.6   \vb 0.64\vb 0.56\vb 0.51\vb 0.47    \cr
  Ideal case\vb 0.98\vb 0.98\vb 0.97\vb 0.97    \endtable
\endinsert
\endpage
\centerline{FIGURE CAPTIONS}
\vskip 1cm
   \item{Fig.1} Distribution of the coupling matrix elements $v_n$ for
             $\chi=1.0$ at three values of $k$. The smooth curves
             show a normalized gaussian distribution having the
             same width as that of $S(E)$ at each $k$ value.

  \item{Fig.2} Strength function $S(E)$ for $\chi=1.0$ at
              $k=$0.0 and 0.6.

  \item{Fig.3} Upper part: Strength distribution $P(S)$ for
                      $\chi=1.0$ at three values of $k$. The smooth
                      curves show the Porter-Thomas distribution.
                Lower part: Distribution $P(\sqrt{\tilde S})$ where
                      ${\tilde S}=S((E-\epsilon)^2+(\Gamma/2)^2)$.
                      $\Gamma$ is fixed to $4\pi$. The smooth curves
                      show the gaussian distribution.

  \item{Fig.4}$M_m(\delta E)$ versus $\delta E$ for $\chi=1.0$
             at $k=$0.0 and 0.6 as well as for the ideal case
             of equidistant energies with equal strengths.
             The lines correspond to $m=$2 to 5 from the upper
             to the lower ones. The arrows indicate the interval
             of $\delta E$ where the fractal dimensions are evaluated.

\bye